\documentclass{article}
\usepackage{spconf,amsmath,graphicx}
\usepackage{hyperref}
\usepackage{bbm}
\usepackage{todonotes}
\usepackage{multirow}
\usepackage{tabularx}
\usepackage{subfig}
\usepackage{booktabs}
\usepackage{enumitem}

\usepackage[misc,geometry]{ifsym}
\usepackage{amsmath,amssymb}
\usepackage{mathtools}
\usepackage{bbold}
\usepackage{mathrsfs}
\usepackage{multirow}
\usepackage{array}
\usepackage{makecell}
\usepackage{tikz}
\usepackage{lipsum}
\usepackage{todonotes}
\usetikzlibrary{arrows.meta, spy, calc, arrows, patterns, patterns.meta}
\usepackage{adjustbox}
\usepackage{color}

\setlist{nosep, leftmargin=14pt}
\usepackage{mwe}

\DeclareMathOperator{\conv}{conv}

\definecolor{pureorange}{RGB}{247,127, 0}

\title{Deep vessel segmentation with joint multi-prior encoding}

\name{A. Sadikine $^{\bullet, \star}$
\qquad \hspace{-0.7cm} B. Badic $^{\bullet, \flat}$
\qquad \hspace{-0.7cm} E. Ferrante $^{\circ}$
\qquad \hspace{-0.7cm} V. Noblet $^{\sharp}$
\qquad \hspace{-0.7cm} P. Ballet $^{\bullet, \star}$
\qquad \hspace{-0.7cm} D. Visvikis $^{\bullet}$
\qquad \hspace{-0.7cm} P.-H. Conze $^{\bullet, \diamond}$
\thanks{This work was partially funded by LaBeX CAMI (grant ANR-11-LABX-0004) and France Life Imaging (grant ANR-11-INBS-0006).}}

\address{$^{\bullet}$ LaTIM UMR 1101, Inserm, Brest, France 
$^{\star}$ University of Western Brittany, Brest, France \\
$^{\flat}$ University Hospital of Brest, Brest, France
$^{\circ}$ CONICET, Santa Fe, Argentina \\ 
$^{\sharp}$ ICube UMR 7357, CNRS, Strasbourg, France
$^{\diamond}$ IMT Atlantique, Brest, France
}

\begin{document}
\maketitle

\begin{abstract}
The precise delineation of blood vessels in medical images is critical for many clinical applications, including pathology detection and surgical planning. However, fully-automated vascular segmentation is challenging because of the variability in shape, size, and topology. Manual segmentation remains the gold standard but is time-consuming, subjective, and impractical for large-scale studies. Hence, there is a need for automatic and reliable segmentation methods that can accurately detect blood vessels from medical images. The integration of shape and topological priors into vessel segmentation models has been shown to improve segmentation accuracy by offering contextual information about the shape of the blood vessels and their spatial relationships within the vascular tree. To further improve anatomical consistency, we propose a new joint prior encoding mechanism which incorporates both shape and topology in a single latent space. The effectiveness of our method is demonstrated on the publicly available 3D-IRCADb dataset. More globally, the proposed approach holds promise in overcoming the challenges associated with automatic vessel delineation and has the potential to advance the field of deep priors encoding. 
\end{abstract}


\section{Introduction}
\label{sec:introduction}


Vessel segmentation is a vital component of medical image analysis, focusing on the precise identification and differentiation of blood vessels within medical images such as Computed Tomography (CT) scans. It holds great significance in various medical applications, including diagnosis, surgical planning, and disease monitoring \cite{madrahimov2006marginal, marvcan2015effect}. However, this task is riddled with challenges including low contrast with surrounding tissues, intricated multi-scale geometry \cite{sadikine2023scale}, and variability in vessel structure. Preserving anatomical features is critical for accurate analysis and treatment planning, as vessel shape and topology provide valuable insight into vessel health and potential disease risks. Manual segmentation is generally tedious, time-consuming and subject to inter- and intra-expert variability, while automated vessel delineation provides a faster and more reliable solution for clinicians, ensuring that essential vessel features are captured.

\begin{figure}[!t]
    \centering
    \includegraphics[width=\linewidth]{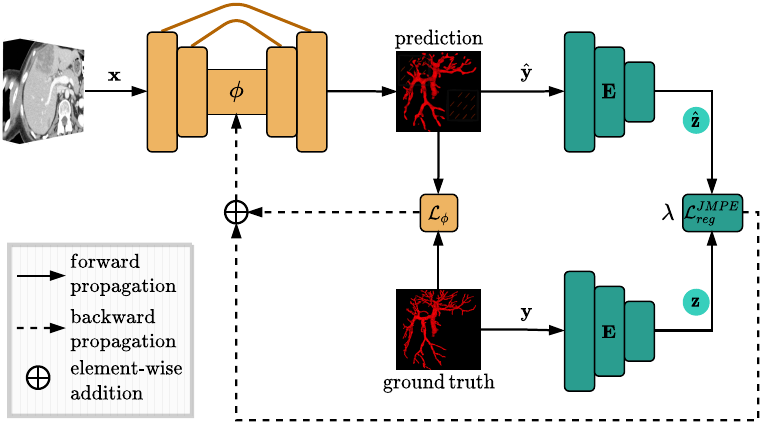} \vspace{-0.5cm} \\
    \caption{Proposed pipeline overview. Parameters of the segmentation model $\phi$ are estimated by penalizing the segmentation loss $\ell_{\phi}$ with a regularization term $\mathcal{L}_{reg}^{JMPE}$ that deals with the similarity between the projections of the prediction $\hat{\pmb{y}}$ and the ground truth $\pmb{y}$ in a learned multi-priors embedding.} \vspace{-0.2cm}
    \label{fig:pipeline}
\end{figure}
The UNet architecture \cite{ronneberger2015u} is a widely recognized Convolutional Neural Network (CNN) used as baseline for image segmentation in medical imaging. In response to the growing complexity of delineation tasks, a multi-task deep learning architecture was introduced in \cite{keshwani2020topnet} for the reconstruction and labeling of hepatic vessels from contrast-enhanced CT scans. The network was able to simultaneously detect vessel centerline voxels and estimate their connectivity, taking into account inter- and intra-class distances between center-voxel pairs. Furthermore, the challenge of complex multi-scale vessel geometry was addressed  in \cite{sadikine2023scale} by introducing a novel deep supervised approach. This method employed a clustering technique to decompose the vascular tree into different scale levels and extended the 3D UNet with multi-task and contrastive learning to enhance inter-scale discrimination.

Despite their success, current segmentation networks can still produce anatomically aberrant vessel segmentations. Recent research has highlighted the importance of incorporating high-level prior knowledge to ensure anatomically plausible delineations \cite{oktay2017anatomically,el2021high,boutillon2022multi}. Such approach improves deep networks by providing additional information during training to capture relevant features from images \cite{conze2023current}. Moreover, anatomical constraints can be introduced during post-processing to refine the contours obtained at inference. Building on this idea, denoising auto-encoders were proposed in \cite{larrazabal2019anatomical} to impose shape constraints on chest X-ray segmentation results. In contrast, shape priors from a Semi-Overcomplete Convolutional Auto-Encoder (S-OCAE) embedding were integrated into deep segmentation networks during the learning process \cite{sadikine2022semi, sadikine2024improving}. While these are well-established approaches, simultaneously incorporating multiple priors in medical imaging segmentation has not received much attention to date.



Incorporating multiple anatomical prior-based loss functions into the segmentation pipeline typically requires the use of multiple individual non-linear encodings. This approach can be problematic because it requires training multiple auto-encoders and tuning multiple hyper-parameters for the prior penalty terms in the loss function. In addition, it can lead to higher memory consumption during training. To address all these drawbacks, we propose a novel approach that involves learning both shape and topological connectivity priors within a unified manifold (Fig.\ref{fig:pipeline}). This is achieved by Joint Multi-Prior Encoding (JMPE), which employs a convolutional encoder derived from a multi-task Convolutional Auto-Encoder (CAE). Its effectiveness is demonstrated for hepatic vessel segmentation on the publicly available 3D-IRCADb dataset.

\section{Methods}
\label{sec:methods}

Let us consider $\pmb{x}$ as a grayscale volume and $\pmb{y}$ its corresponding binary ground truth segmentation. Deep supervised segmentation involves formulating a mapping function $\phi:\pmb{x}\rightarrow \hat{\pmb{y}}$. This mapping function is optimized through the optimization of a loss function $\mathcal{L}_{\phi}(\pmb{y},\hat{\pmb{y}})$, which in our case consists of a combination of both weighted binary cross-entropy and Dice loss components. However, such loss functions are defined at the pixel level and do not have the abiltiy to account for high-level features or topological characteristics of the target. In this context, our work focuses on the process of integrating multiple priors into the segmentation pipeline, through a compact and non-linear encoding.

\subsection{Shape and topology information}

The geometric and spatial characteristics of tubular structures are crucial for discerning their shape and overall topology. Cross-sectional radii and skeleton primarily characterize their geometrical properties. Determining the shape of a given vessel tree is typically performed by medical experts according to their domain knowledge. This process leverages spatial coordinates and prior knowledge of the anatomy to create a ground truth segmentation mask $\pmb{y}$. In binary scenarios, this segmentation mask is defined as:


\begin{equation}
    \pmb{y}_{i}(\nu) =
    \begin{cases}
        1 \quad \text{if} \; \nu\in \mathcal{S}\\
        0\quad \text{otherwise}
    \end{cases}
    \label{binary-mask}
\end{equation} \vspace{-0.3cm} \\

\noindent where $\nu$ is the set of voxels belonging to $\pmb{y}_{i}$ and $\mathcal{S}$ the spatial domain of the vascular structure of interest.

In addition, topological connectivity refers to the arrangement and connection of components within the targeted vascular structure. It involves analyzing how different parts of the structure are connected, including branching points, bifurcations, and endpoints. This connectivity can be effectively captured by abstract representations through skeletonization. Alternatively, the Euclidean Distance Transform (EDT) \cite{rosenfeld1968distance} provides another approach to encode this topological property within tubular structures into a distance map, noted as $T_i$, where ridge points \cite{ge1996generation} correspond to the skeleton of the EDT. This representation ensures a seamless continuity between the ridge-based skeleton and its adjacent voxels. The EDT is a well-known technique for computing the minimum Euclidean distance between each voxel and the nearest background voxel surface $\Omega$. This calculation is expressed as:

\begin{equation}
T_i(\nu)= 
    \begin{cases} 
    \min_{\upsilon \in \Omega} \|\nu-\upsilon\|_2 &  \text{if } \nu \in \mathcal{S} \\ 
    0 &   \text {otherwise}
    \end{cases}
\end{equation}  \vspace{-0.3cm} \\

To extract high-level features for shape and topology from mask $\pmb{y}_{i}$ and distance map $T_i$, the process involves defining the transformation that represents both shape and topological connectivity into a compact non-linear representation.

\subsection{Deep prior encoding}

Towards deep prior incorporation, we are interested in learning a mapping function that captures high-level information from observations $\pmb{a}_i$. This mapping function, denoted as $E_{\theta}$ and parameterized by $\theta$, transforms the input data $\pmb{a}$ into a high-level undercomplete summary represented as $\pmb{z}$, with a smaller size compared to the input, as this design choice allows to capture global information in a compact form. On the other hand, the decoder function $D$ maps the latent code $\pmb{z}$ back to the original observation space, generating an approximate reconstruction $\tilde{\pmb{a}}$. The entire process is characterized by the pair of functions $E_{\theta}: \mathcal{A} \rightarrow \mathcal{Z}$ and $D: \mathcal{Z} \rightarrow \mathcal{A}$. In essence, $D$ is applied to the latent representation $\pmb{z}$ obtained from $E_{\theta}$, resulting in the reconstructed data $\tilde{\pmb{a}}=D\circ E_{\theta}(\pmb{a})$. The parameters of such convolutional auto-encoder architecture are estimated by minimizing the following loss function:

\begin{equation}\label{caeloss}
    \mathcal{L}_{CAE}(\pmb{a},\tilde{\pmb{a}})\propto \sum_{a}{\ell_i(\pmb{a}_i, \tilde{\pmb{a}}_i)}
\end{equation} \vspace{-0.15cm}

\noindent where $\ell_i$ is the individual loss for each data sample $\pmb{a}_i$, computed as the weighted average of smooth L1 loss values:

\begin{equation}
\ell_i = \frac{1}{N}  \sum_{j, k, m=1} w_{jkm} \cdot \texttt{smooth}_{L1}(\pmb{a}_i(j,k,m), \tilde{\pmb{a}}_i(j,k,m))
\end{equation}

\noindent For a given coordinate $(j,k,m)$, the weight is set as follows:

\begin{equation}
w_{jkm}=\frac{N}{ \begin{cases}N_{pos}, & \text{if} \;  \pmb{a}_i(j,k,m)=1 \\ N_{neg}, &\text{if} \;  \pmb{a}_i(j,k,m)=0 \end{cases} }
\end{equation} \vspace{0.1cm}

\noindent Here, $N$ is the total count of voxels in $\pmb{y}_i$. $N_{pos}$ and $N_{neg}$ represent the count of positive and negative voxels, respectively. In the context of shape encoding, the input is represented as $\pmb{a}_i$, which is defined as $\pmb{y}_i$, and the output is designated as $\tilde{\pmb{a}}_i$, mirroring the reconstruction of $\pmb{y}_i$. In contrast, when encoding distance maps $T_i$, the input is expressed with respect to $\pmb{y}_i$, and the output ($\Tilde{T}_i$) tends to align with the generated distance map, following a regression problem formulation.

We can measure anatomical alignment \cite{oktay2017anatomically}, focusing on shape or topology, by comparing ground truth to the segmentation model prediction (Fig.\ref{fig:pipeline}). This alignment is assessed through the lower-dimensional representation generated by the learned encoder $E_{\theta}=\pmb{z}^p$, employing a distance $d(.)$:

\begin{equation}\label{regterm}
    \mathcal{L}_{reg}^{p}(\pmb{y},\hat{\pmb{y}}) \propto \sum_{y}{d({E_{\theta}(\pmb{y}), E_{\theta}(\hat{\pmb{y}})})}
\end{equation} \vspace{-0.25cm}

\noindent where $p$ can take two possible values: $p = s$ (shape) or $p = t$ (topology). This choice of $p$ determines the specific type of alignment, whether it relates to shape or topology. The regularization term $\mathcal{L}_{reg}^{p}$ is subsequently added to the segmentation loss during training (Fig.\ref{fig:pipeline}). This addition of the regularization term is essential for incorporating contextual information throughout the training of the segmentation model:

\begin{equation}\label{globalloss1}
   \mathcal{L}_t = \mathcal{L}_{\phi}(\pmb{y},\hat{\pmb{y}}) + \lambda_p\mathcal{L}_{reg}^{p}(\pmb{y},\hat{\pmb{y}})
\end{equation} \vspace{-0.2cm}

\noindent where $\lambda_p$ is an empirically determined hyper-parameter that balances the contribution of the penalty term. However, the loss function (Eq.\ref{globalloss1}) is employed when incorporating either shape or topology priors. In the event of introducing both simultaneously, the modified loss function is given as:

\begin{equation}\label{globalloss2}
   \mathcal{L}_t = \mathcal{L}_{\phi}(\pmb{y},\hat{\pmb{y}}) + \lambda_s\mathcal{L}_{reg}^{s}(\pmb{y},\hat{\pmb{y}}) + \lambda_t\mathcal{L}_{reg}^{t}(\pmb{y},\hat{\pmb{y}})
\end{equation} \vspace{-0.2cm}

Incorporating both shape and topology into the loss function requires training two separate encoders and setting two different hyper-parameters (Eq.\ref{globalloss2}), which can be cumbersome and add complexity to the training process. To overcome this, $\mathcal{L}_{reg}^{s}$ and $\mathcal{L}_{reg}^{t}$ can be combined into an unified formulation.

\begin{figure}[!t]
    \centering
    \includegraphics[width=0.95\linewidth]{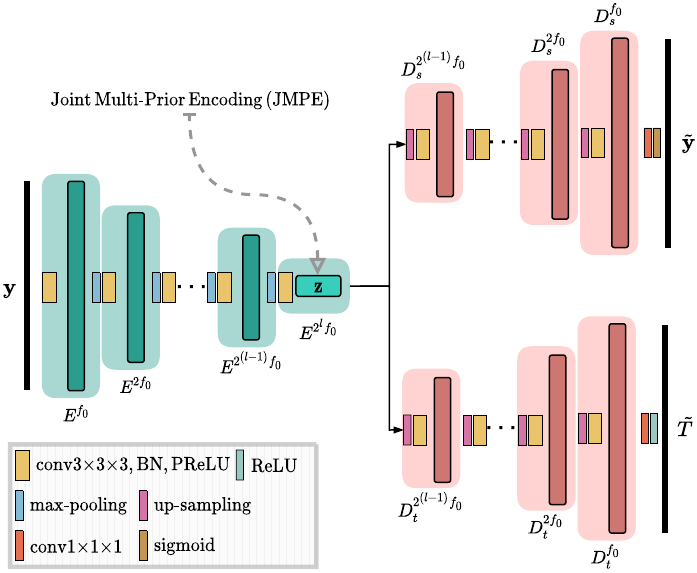} \vspace{-0.1cm} \\
    \caption{Multi-task convolutional auto-encoder network $\xi$ architecture for Joint Multi-Prior Encoding (JMPE).} \vspace{-0.2cm}
    \label{fig:jmpe}
\end{figure}

\begin{table*}[ht!]
\centering
\vspace{-0.25cm}
\caption{Liver vessel CT segmentation on 3D-IRCADb \cite{soler20103d} using 3D ResUNet as baseline. Models incorporating shape and topological constraints, a mix of both, and our own approach are compared. Best results in bold, second best results underlined.}


\resizebox{1.7\columnwidth}{!}{%

\begin{tabular}{@{}llcccccc@{}}\cmidrule(l){3-8} 
\multicolumn{2}{c}{Models} &
  \shortstack{\texttt{DSC}$\uparrow$ \\ score ($\%$)} &
  \shortstack{\texttt{Jacc}$\uparrow$ \\ score ($\%$)} &
  \shortstack{\texttt{clDSC}$\uparrow$ \\ score ($\%$)} &
  \shortstack{\texttt{HD}$\downarrow$\\ dist. ($\mathrm{mm}$)} &
  \shortstack{\texttt{AVD}$\downarrow$\\ dist. ($\mathrm{mm^3}$)} &
  \shortstack{\texttt{ASSD}$\downarrow$\\ dist. ($\mathrm{mm}$)} \\ \midrule
ResUNet   & \hfil -          & $54.06\pm2.34$       & $37.31\pm2.14$        & $46.31\pm2.37$       & $70.88\pm9.20 $ &$0.36\pm0.18$ & $5.23\pm0.75$\\
ResUNet+shape                              & \hfil $\lambda_{s}=26.21$     & $53.50\pm2.44$       & $36.97\pm2.31$       & $44.50\pm3.03$       & $\underline{61.88}\pm5.02$  &$0.35\pm0.18$ & $5.20\pm0.69$\\   
ResUNet+topo                                & \hfil $\lambda_{t}=32.01$      & $53.50\pm1.50$       & $36.84\pm1.40$       & $47.41\pm1.81$       & $\pmb{61.72}\pm5.74$   &$0.37\pm0.15$ & $5.09\pm0.73$\\  
ResUNet+shape+topo & $\hfil \lambda_{s}=63.33, \lambda_{t}=14.53$   & $\underline{54.59}\pm2.06$       & $\underline{37.88}\pm1.87$       & $\underline{47.63}\pm2.72$       & $71.50\pm8.44$       &$\pmb{0.33}\pm0.15$ & $\underline{4.81}\pm0.72$\\
    \textbf{Ours}                           & \hfil $\lambda=65.10$ & $\pmb{54.78}\pm2.35$       & $\pmb{38.00}\pm2.17$      & $\pmb{50.34}\pm2.84$       & $67.06\pm5.27$      &$\underline{0.34}\pm0.15$ & $\pmb{4.77}\pm0.74$\\ \midrule
                               &                 & \multicolumn{1}{l}{} & \multicolumn{1}{l}{} & \multicolumn{1}{l}{} & \multicolumn{1}{l}{}
\end{tabular}%
}

\vspace{-0.6cm}
\label{tab:qresults}
\end{table*}

\subsection{Proposed deep joint multi-prior encoding}

The pursuit of learning multiple priors in a unified compact representation $\pmb{z}$, which we refer to as Joint Multi-Prior Encoding (JMPE), stands as a more efficient alternative than employing separate encodings $\pmb{z}^p$. This challenge is effectively addressed through a multi-task learning approach, facilitated by a single encoder $E_{\theta}$ and multiple decoders $D_p$, all sharing the same latent code representation $\pmb{z}$. This technique proves particularly valuable in applications where various tasks exhibit inter-dependencies, offering a streamlined and holistic approach to jointly managing multiple priors representation in a single latent space. The formulation of the multi-task convolutional auto-encoder $\xi$ (Fig.\ref{fig:jmpe}), is defined as: \vspace{-0.25cm}

\begin{equation} 
    \xi(\pmb{y}) \coloneqq \{D_s(\pmb{z})=\tilde{\pmb{y}}, D_t(\pmb{z})=\Tilde{T} \hspace{0.1cm}|\hspace{0.1cm} \pmb{z}=E_{\theta}(\pmb{y})\}
\end{equation} \vspace{-0.25cm}

\noindent where $D_s$ and $D_t$ are dedicated to the tasks of reconstruction and regression, respectively. The optimal model $\xi$ is achieved by minimizing the following loss across all training tasks:
 
\begin{equation}\label{jmpeloss}
    \mathcal{L}_{JMPE}(\pmb{y},\tilde{\pmb{y}})\propto \alpha_s\sum_{y}{\ell_i(\pmb{y}_i, \tilde{\pmb{y}}_i)} + \alpha_t\sum_{T_i}{\ell_i(T_i, \Tilde{T}}_i)
\end{equation} \vspace{-0.05cm}

\noindent where $\alpha_p$ weighting factors aim at balancing both tasks during training. Once the network $\xi$ has been trained, the anatomical alignment $\mathcal{L}_{reg}^{JMPE}$  can be computed analogously to that shown in Eq.\ref{regterm}. This is achieved by quantifying the distance between the projections of $\pmb{y}$ and $\hat{\pmb{y}}$. The encoder $E_{\theta}$, followed by a ${\conv}1{\times}1{\times}1$ operation with fixed weights, is used to reduce the number of latent code feature maps, thereby improving the capture of more subtle features. In this scenario, Eq.\ref{globalloss2} is streamlined into a unified regularization term:

\begin{equation}\label{globalloss3}
   \mathcal{L}_t = \mathcal{L}_{\phi}(\pmb{y},\hat{\pmb{y}}) + \lambda\mathcal{L}_{reg}^{JMPE}(\pmb{y},\hat{\pmb{y}})
\end{equation} \vspace{-0.2cm}

\section{Experiments}
\label{sec:experiment}

\subsection{Imaging datasets}

The 3D-IRCADb \cite{soler20103d} dataset includes contrast-enhanced CT scans from 20 patients, equally divided between 10 females and 10 males. In approximately $75\%$ of cases, these scans show the presence of liver tumors. Expert radiologists manually annotated ground truth masks for the liver, liver vessels, and liver tumors. Pre-processing included resampling to a median voxel spacing, cropping to focus on the liver area, and appropriate clipping ([-150, 250]) of CT intensities.

\subsection{Implementation details}

Throughout the encoding stage, we set the following parameters: the number of layers $l$ to $5$, the initial number of feature maps $f_0$ to $8$ (Fig.\ref{fig:jmpe}), $\alpha_p$ in Eq.\ref{jmpeloss} to $1$, the number of latent code feature maps to 32, the learning rate to $10^{-4}$, the batch size to $2$, and the number of epochs to $1000$. In contrast, for the segmentation experiments, the learning rate, batch size, and number of epochs were set to $3\times10^{-4}$, 2, and 1500, respectively. Additionally, the distance function $d(\cdot)$ was defined as the cosine distance (Eq.\ref{regterm}). Hyper-parameter optimization was performed using Optuna \cite{optuna_2019} with 20 trials for each configuration, and optimal $\lambda$ values were determined empirically as shown in Tab.\ref{tab:qresults}. Random data augmentation techniques including rotation, translation, flipping, and gamma correction was applied. A 5-fold cross-validation approach was used. Deep networks were implemented using PyTorch. In practice, seeds were fixed for weight initialization, data augmentation and shuffling to ensure reproducibility.

\begin{figure}[!t]
\begin{tabular}{ccc}
\hspace{0.025cm} \footnotesize ground truth &
\hspace{-0.4cm} \footnotesize ResUNet &
\hspace{-0.4cm} \footnotesize ResUNet+shape \cr
\hspace{0.025cm} \includegraphics[width=2.6cm]{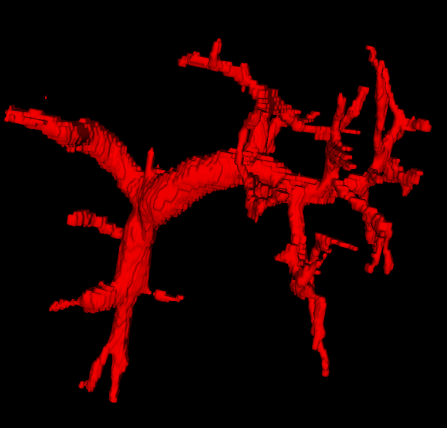} &
\hspace{-0.4cm} \includegraphics[width=2.6cm]{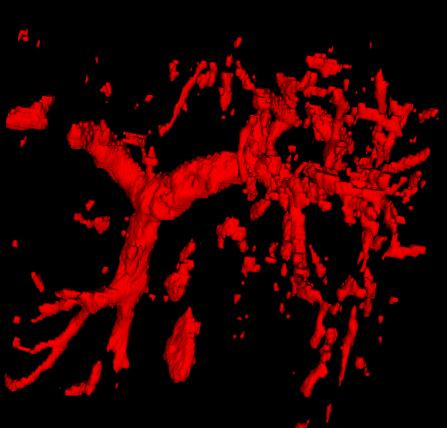} &
\hspace{-0.4cm} \includegraphics[width=2.6cm]{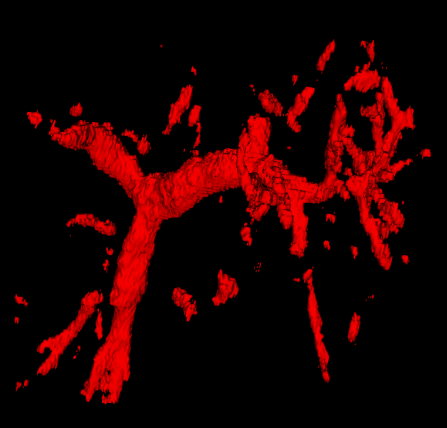} \vspace{-0.1cm} \cr
\hspace{0.025cm} \footnotesize ResUNet+topo &
\hspace{-0.4cm} \footnotesize ResUNet+sh.+topo &
\hspace{-0.4cm} \footnotesize \textbf{Ours} \cr
\hspace{0.025cm} \includegraphics[width=2.6cm]{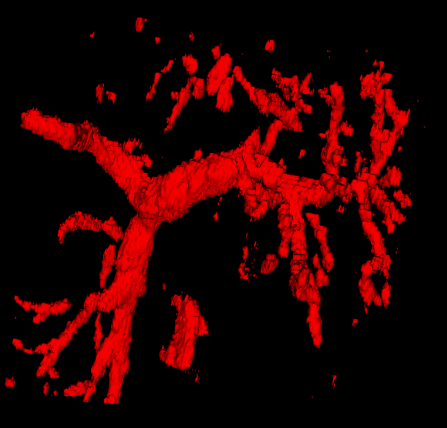} &
\hspace{-0.4cm} \includegraphics[width=2.6cm]{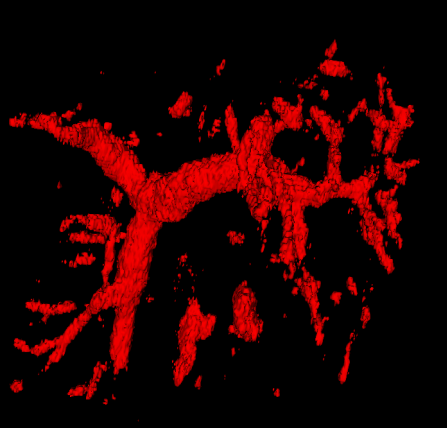} &
\hspace{-0.4cm} \includegraphics[width=2.6cm]{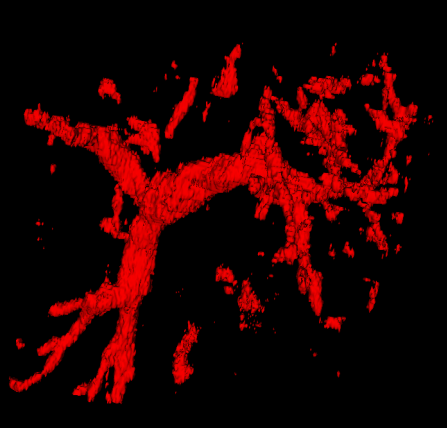} \cr
\end{tabular}
\vspace{-0.7cm} \\
\caption{Liver vessel segmentation results on 3D-IRCADb \cite{soler20103d} using various priors and 3D ResUNet as backbone.} \vspace{-0.2cm}
\label{fig:qualres}
\end{figure} 



\section{Results and discussion}
\label{sec:results}

Results in Tab.\ref{tab:qresults} show the performance of different models for liver vessel segmentation, assessed using various evaluation metrics including \texttt{DSC} (Dice Similarity Coefficient), \texttt{Jacc} (Jaccard score), \texttt{clDSC} coefficient \cite{shit2021cldice} for connectivity assessment, \texttt{HD} (Hausdorff distance), \texttt{AVD} (Absolute Volume Difference), and \texttt{ASSD} (Average Symmetric Surface Distance). The models include ResUNet as baseline, ResUNet+shape, ResUNet+topo, ResUNet+shape+topo, and the proposed approch. Our method outperforms the other models, achieving the highest \texttt{DSC}, \texttt{Jacc}, \texttt{clDSC}, and \texttt{ASSD} scores with $54.78\%$, $38.00\%$, $50.34\%$, and $4.77$mm respectively. It delivers robust performance in \texttt{clDSC} assessment, positioning it as a promising topology-aware model. Further, Fig.\ref{fig:qualres} illustrates the connectivity improvement reached by our approach. In particular, fine branches remain less disconnected from main vessels compared to ResUNet+topo or ResUNet+shape+topo. The performance of our approach could be improved by calibrating the hyper-parameter $\alpha^p$ (Eq.\ref{jmpeloss}), which indirectly affects the JMPE coding scheme. Furthermore, our method allows the use of a single encoder instead of multiple encoders, thus reducing memory consumption.

\section{Conclusion}
\label{sec:conclusion}

In this paper, we presented an innovative approach that addresses the integration of multiple priors into a unified formulation for segmentation purposes. The liver vessel delineation results obtained from our method highlight the importance of incorporating high-level and topological constraints in medical image segmentation, and provide potential avenues for future research in this area. Furthermore, extending this approach to other datasets could provide valuable insights into its generalizability and effectiveness in various clinical contexts. Additionally, integrating graph neural networks in our pipeline could further enhance connectivity encoding.


\section{Compliance with ethical standards}
This research study was conducted retrospectively using human subject data made available in open access \cite{soler20103d}. The authors declare that they do not have any conflicts of interest.

\bibliographystyle{IEEEbib}
\bibliography{References}

\begin{thebibliography}{10}

\bibitem{madrahimov2006marginal}
Nodir Madrahimov, Olaf Dirsch, Christoph Broelsch, and Uta Dahmen,
\newblock ``Marginal hepatectomy in the rat: from anatomy to surgery,''
\newblock {\em Annals of Surgery}, vol. 244, no. 1, pp. 89, 2006.

\bibitem{marvcan2015effect}
Marija Mar{\v{c}}an, Bor Kos, and Damijan Miklav{\v{c}}i{\v{c}},
\newblock ``Effect of blood vessel segmentation on the outcome of electroporation-based treatments of liver tumors,''
\newblock {\em PLoS One}, vol. 10, no. 5, pp. e0125591, 2015.

\bibitem{sadikine2023scale}
Amine Sadikine, Bogdan Badic, Jean-Pierre Tasu, Vincent Noblet, Pascal Ballet, Dimitris Visvikis, and Pierre-Henri Conze,
\newblock ``Scale-specific auxiliary multi-task contrastive learning for deep liver vessel segmentation,''
\newblock in {\em IEEE International Symposium on Biomedical Imaging}, 2023.

\bibitem{ronneberger2015u}
Olaf Ronneberger, Philipp Fischer, and Thomas Brox,
\newblock ``U{N}et: Convolutional networks for biomedical image segmentation,''
\newblock in {\em Medical Image Computing and Computer-Assisted Intervention}, 2015, pp. 234--241.

\bibitem{keshwani2020topnet}
Deepak Keshwani, Yoshiro Kitamura, Satoshi Ihara, Satoshi Iizuka, and Edgar Simo-Serra,
\newblock ``Top{N}et: Topology preserving metric learning for vessel tree reconstruction and labelling,''
\newblock in {\em Medical Image Computing and Computer Assisted Intervention}, 2020, pp. 14--23.

\bibitem{oktay2017anatomically}
Ozan Oktay, Enzo Ferrante, Konstantinos Kamnitsas, Mattias Heinrich, Wenjia Bai, Jose Caballero, Stuart~A Cook, Antonio De~Marvao, Timothy Dawes, Declan~P O‘Regan, et~al.,
\newblock ``Anatomically constrained neural networks ({ACNN}s): application to cardiac image enhancement and segmentation,''
\newblock {\em IEEE Transactions on Medical Imaging}, vol. 37, no. 2, pp. 384--395, 2017.

\bibitem{el2021high}
Rosana El~Jurdi, Caroline Petitjean, Paul Honeine, Veronika Cheplygina, and Fahed Abdallah,
\newblock ``High-level prior-based loss functions for medical image segmentation: {A} survey,''
\newblock {\em Computer Vision and Image Understanding}, vol. 210, pp. 103248, 2021.

\bibitem{boutillon2022multi}
Arnaud Boutillon, Bhushan Borotikar, Val{\'e}rie Burdin, and Pierre-Henri Conze,
\newblock ``Multi-structure bone segmentation in pediatric {MR} images with combined regularization from shape priors and adversarial network,''
\newblock {\em Artificial Intelligence in Medicine}, vol. 132, pp. 102364, 2022.

\bibitem{conze2023current}
Pierre-Henri Conze, Gustavo Andrade-Miranda, Vivek~Kumar Singh, Vincent Jaouen, and Dimitris Visvikis,
\newblock ``Current and emerging trends in medical image segmentation with deep learning,''
\newblock {\em IEEE Transactions on Radiation and Plasma Medical Sciences}, 2023.

\bibitem{larrazabal2019anatomical}
Agostina~J Larrazabal, Cesar Martinez, and Enzo Ferrante,
\newblock ``Anatomical priors for image segmentation via post-processing with denoising autoencoders,''
\newblock in {\em Medical Image Computing and Computer Assisted Intervention}, 2019, pp. 585--593.

\bibitem{sadikine2022semi}
Amine Sadikine, Bogdan Badic, Jean-Pierre Tasu, Vincent Noblet, Dimitris Visvikis, and Pierre-Henri Conze,
\newblock ``Semi-overcomplete convolutional auto-encoder embedding as shape priors for deep vessel segmentation,''
\newblock in {\em IEEE International Conference on Image Processing}, 2022, pp. 586--590.

\bibitem{sadikine2024improving}
Amine Sadikine, Bogdan Badic, Jean-Pierre Tasu, Vincent Noblet, Pascal Ballet, Dimitris Visvikis, and Pierre-Henri Conze,
\newblock ``Improving abdominal image segmentation with overcomplete shape priors,''
\newblock {\em Computerized Medical Imaging and Graphics}, vol. 113, pp. 102356, 2024.

\bibitem{rosenfeld1968distance}
Azriel Rosenfeld and John~L Pfaltz,
\newblock ``Distance functions on digital pictures,''
\newblock {\em Pattern Recognition}, vol. 1, no. 1, pp. 33--61, 1968.

\bibitem{ge1996generation}
Yaorong Ge and J~Michael Fitzpatrick,
\newblock ``On the generation of skeletons from discrete {E}uclidean distance maps,''
\newblock {\em IEEE Transactions on Pattern Analysis and Machine Intelligence}, vol. 18, no. 11, pp. 1055--1066, 1996.

\bibitem{soler20103d}
L~Soler, A~Hostettler, V~Agnus, A~Charnoz, J~Fasquel, J~Moreau, A~Osswald, M~Bouhadjar, and J~Marescaux,
\newblock ``{3D} image reconstruction for comparison of algorithm database: A patient specific anatomical and medical image database,''
\newblock {\em IRCAD Tech. Rep}, 2010.

\bibitem{optuna_2019}
Takuya Akiba, Shotaro Sano, Toshihiko Yanase, Takeru Ohta, and Masanori Koyama,
\newblock ``Optuna: A next-generation hyperparameter optimization framework,''
\newblock in {\em International Conference on Knowledge Discovery and Data Mining}, 2019.

\bibitem{shit2021cldice}
Suprosanna Shit, Johannes~C Paetzold, Anjany Sekuboyina, Ivan Ezhov, Alexander Unger, Andrey Zhylka, Josien~PW Pluim, Ulrich Bauer, and Bjoern~H Menze,
\newblock ``cl{D}ice - {A} novel topology-preserving loss function for tubular structure segmentation,''
\newblock in {\em IEEE Conference on Computer Vision and Pattern Recognition}, 2021, pp. 16560--16569.

\end{thebibliography}

\end{document}